\documentclass[]{spie}  

\usepackage{amsmath,amsfonts,amssymb}
\usepackage{soul}
\usepackage{subfig}
\usepackage{xcolor}
\usepackage{graphicx}
\usepackage{floatrow}
\usepackage{ctable}
\usepackage{color, colortbl}
\usepackage{multirow}
\usepackage[colorlinks=true,linkcolor=blue, citecolor=blue, urlcolor=blue]{hyperref}
\definecolor{newcolor}{rgb}{.8,.349,.1}
\definecolor{Gray1}{gray}{0.94}
\definecolor{Gray2}{gray}{0.84}
\definecolor{applegreen}{rgb}{0.55, 0.71, 0.0}

\usepackage{times}
\title{Deep Angiogram: Trivializing Retinal Vessel Segmentation}

\author[a]{Dewei Hu}
\author[b]{Xing Yao}
\author[b]{Jiacheng Wang}
\author[c]{Yuankai K. Tao}
\author[b]{Ipek Oguz}
\affil[a]{Vanderbilt University, Dept. of Electrical and Computer Engineering}
\affil[b]{Vanderbilt University, Dept. of Computer Science}
\affil[c]{Vanderbilt University, Dept. of Biomedical Engineering, Nashville, TN, USA}

\authorinfo{Send correspondence to Ipek Oguz (ipek.oguz@vanderbilt.edu)}

\begin{document} 
\maketitle

\begin{abstract}

Among the research efforts to segment the retinal vasculature from fundus images, deep learning models consistently achieve superior performance. However, this data-driven approach is very sensitive to domain shifts. For fundus images, such data distribution changes can easily be caused by variations in illumination conditions as well as the presence of disease-related features such as hemorrhages and drusen. Since the source domain may not include all possible types of pathological cases, a model that can robustly recognize vessels on unseen domains is desirable but remains elusive, despite many proposed segmentation networks of ever-increasing complexity. In this work, we propose a contrastive variational auto-encoder that can filter out irrelevant features and synthesize a latent image, named deep angiogram, representing only the retinal vessels. Then segmentation can be readily accomplished by thresholding the deep angiogram. The generalizability of the synthetic network is improved by the contrastive loss that makes the model less sensitive to variations of image contrast and noisy features. Compared to baseline deep segmentation networks, our model achieves higher segmentation performance via simple thresholding. Our experiments show that the model can generate stable angiograms on different target domains, providing excellent visualization of vessels and a non-invasive, safe alternative to fluorescein angiography. 

\end{abstract}


\section{INTRODUCTION}
\label{sec:intro} 
Retinal fundus photography is a cheap, fast and non-invasive modality that reveals essential anatomical features including optic disc, optic cup, macula, fovea, vessels and lesions such as hemorrhages and exudates~\cite{li2021applications}. Therefore, it is widely used for the diagnosis of diseases such as diabetic retinopathy~\cite{islam2020deep}, glaucoma~\cite{zhang2010origa} and age-related macular degeneration~\cite{spaide2003fundus}. 
While fundus photography is broadly used as a low-cost screening tool, it does not provide sufficient contrast to resolve clinically relevant vascular features and exogenous indocyanine green angiography (ICG)/fluorescein angiography (FA) remain the standard of care for visualization/quantifying retinal vasculopathies. An algorithm that can provide accurate vessel segmentation from these fundus images would have profound impact on future clinical practice. In recent years, deep learning  models ~\cite{chen2021retinal} have achieved remarkable success in this task. Nevertheless, the domain shift induced by variations in image contrast and presence of unseen pathological features in testing data can dramatically degrade the performance of deep models.

Recent research explored three main types of domain generalization methods~\cite{wang2022generalizing}: domain randomization, representation learning and general learning strategy. Domain randomization augments the training data to extend the source domain~\cite{zhang2022semi}, improving the likelihood that an unseen target domain overlaps with the training domain. Representation learning refers to the disentanglement of features that are invariant to different domains~\cite{jiang2020unified}. A typical  general learning strategy is meta-learning: for example, Li et al.\ simulate the domain shift by splitting the source domain into meta-train and meta-test~\cite{li2018learning}. 

In this work, we leverage both domain randomization and representation learning approaches to train a model that has superior generalizability across different domains. We augment the source domain by the contrast limited adaptive histogram equalization (CLAHE)~\cite{reza2004realization} with clip limit $\epsilon\in \mathcal{N}$. In addition to well-enhanced contrast for vessels, the augmented images also have exaggerated irrelevant structures including noise and lesions. Inspired by the idea of disentangling the shared features in two images presented in our previous work~\cite{hu2021domain,hu2021life}, we leverage a variational auto-encoder (VAE) to extract the representation of vessels. However, as we showed in~\cite{hu2021domain}, this latent image may have an arbitrary style that contains unwanted features. We tackle this challenge by introducing a contrastive loss such that vessels are the only features in the synthetic image. We name the result a \textit{deep angiogram}. Then, the segmentation task is simply reduced to Otsu thresholding~\cite{otsu1979threshold}. Without the irrelevant features, the visibility of the vasculature is drastically improved in the deep angiogram compared to other vessel enhancement approaches~\cite{subramaniam2022vessel}. We evaluate the generalizability of our model by the segmentation performance on the target domains. For baseline models, we trained two segmentation networks on the source domain that take the green channel fundus image and the principle component analysis (PCA) image as the input respectively. The result indicates that the proposed method generalizes better on target domains and achieves higher segmentation performance than deep segmentation networks, by simple thresholding.

\section{METHODS}

\begin{figure}[t]
    \centering
    \subfloat[\centering model structure]{\includegraphics[width=.55\linewidth]{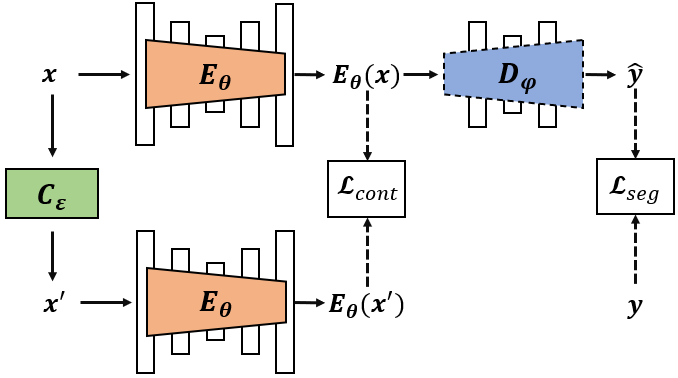}}
    \qquad
    \subfloat[\centering source and target domains]{\includegraphics[width=.37\linewidth]{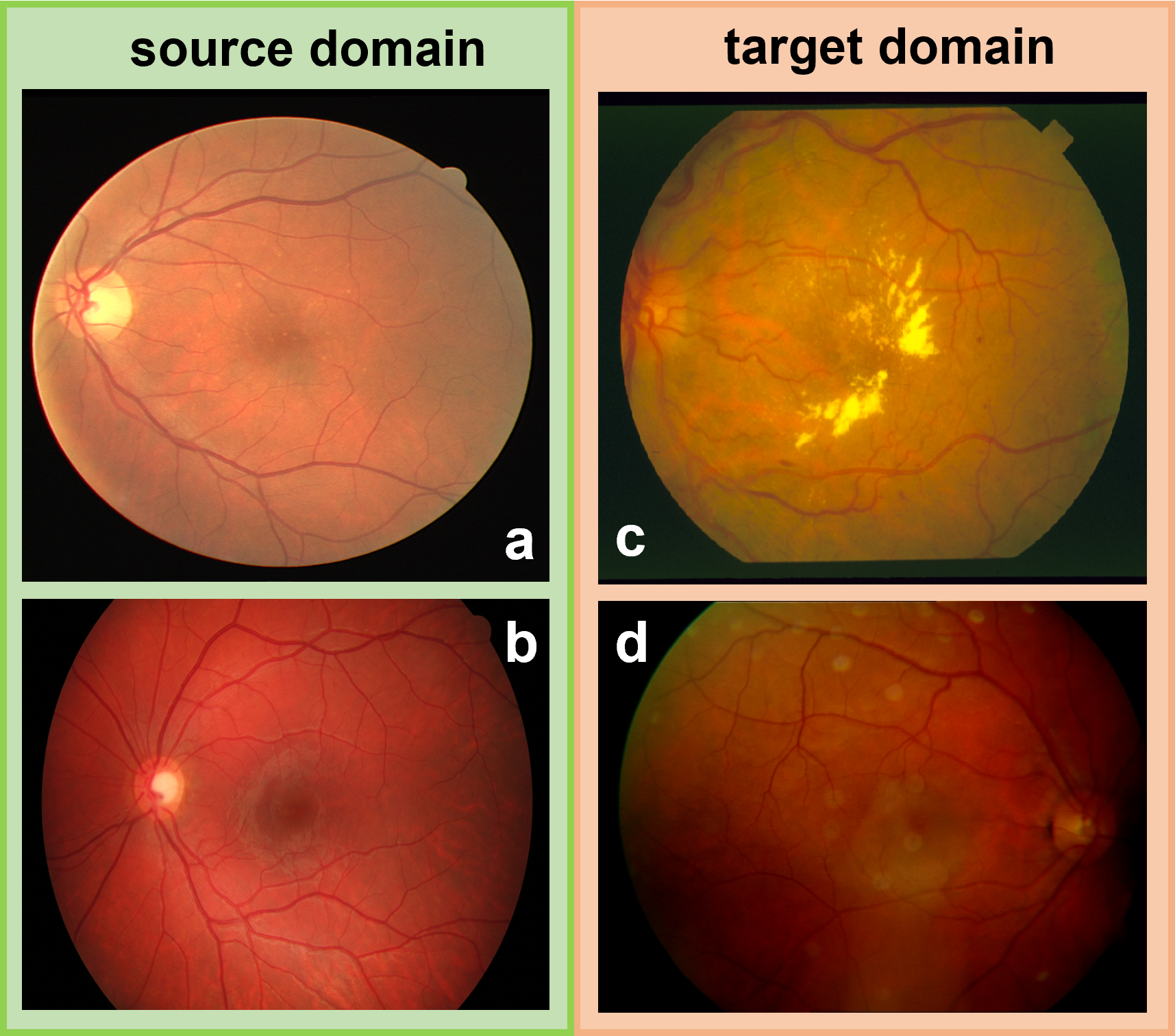}}
    \caption{{\bf (a)} The deep angiogram model structure. $x$ is the input fundus image, $x' = C_{\epsilon}(x)$ is the CLAHE-enhanced image with clip limit $\epsilon$. $y$ is the ground truth. $E_{\theta}$ is a residual U-Net that serves as the encoder of the VAE. $D_\varphi$ is the corresponding decoder. $\mathcal{L}_{cont}$ and $\mathcal{L}_{seg}$ represent the contrastive loss and segmentation loss. The dashed line on $D_\varphi$ indicates it will not be applied in testing. {\bf (b)} The source  target domains.  \textbf{a,} DRIVE, \textbf{b,} HRF, \textbf{c,} STARE, \textbf{d,} ARIA.}
    \label{fig:model}
\end{figure}

\subsection{Causal Feature Extraction}
Fig.~\ref{fig:model}(a) shows our VAE model composed by the encoder $E_{\theta}$ and the decoder $D_{\varphi}$. The input image is $x$ and the supervision is provided by the label $y$. As we have previously shown~\cite{hu2021domain,hu2021life},  when the latent manifold of the VAE has the same dimension with input $x$, the encoder is able to enhance the shared features in $x$ and $y$. Intuitively, if an image is regarded as a collection of representations, then $(x \cap y) \subseteq E_{\theta}(x)$ should hold to guarantee that there is no essential information missing in the output $\hat{y}$. In the context of causal learning, $x \cap y$ is the set of causal features for the final prediction. In this implementation, the fundus image $x$ includes information of many anatomical structures such as optic disc, vessels, macula and lesions, whereas the causal features for the segmentation task contain just the vessels, so ideally the latent image should be a vessel map without any irrelevant features, i.e., $(x \cap y) = E_{\theta}(x)$.

As suggested in Fig.~\ref{fig:model}, since we want to put most of the workload on the encoder $E_{\theta}$, it is designed to have more learnable parameters than the decoder $D_{\varphi}$. Both $E_{\theta}$ and $D_{\varphi}$ have residual U-Net architecture. Note that the decoder $D_{\varphi}$ will not be applied in the testing since its purpose is to simply provide supervision to $E_{\theta}$ during training. The segmentation loss for the decoder is set to be a combination of cross-entropy and Dice loss:
\begin{equation}
    \mathcal{L}_{seg}=-\frac{1}{N}\sum_{n=1}^Ny_n\log\hat{y}_n + \left(1-\frac{2\sum_{n=1}^N y_n\hat{y}_n}{\sum_{n=1}^N y_{n}^{2}+\hat{y}_n^2}\right)
\end{equation}

\newcommand{\imwidth}{0.31}
\begin{figure}[t]
    \centering
    \begin{tabular}{c@{}c@{}c@{}c}
         \textbf{Fundus} & \textbf{Deep Angiogram} & \textbf{Label} \\
         \rotatebox{90}{\hspace{3.0cm}\textbf{ARIA}}
         \includegraphics[width=\imwidth\linewidth]{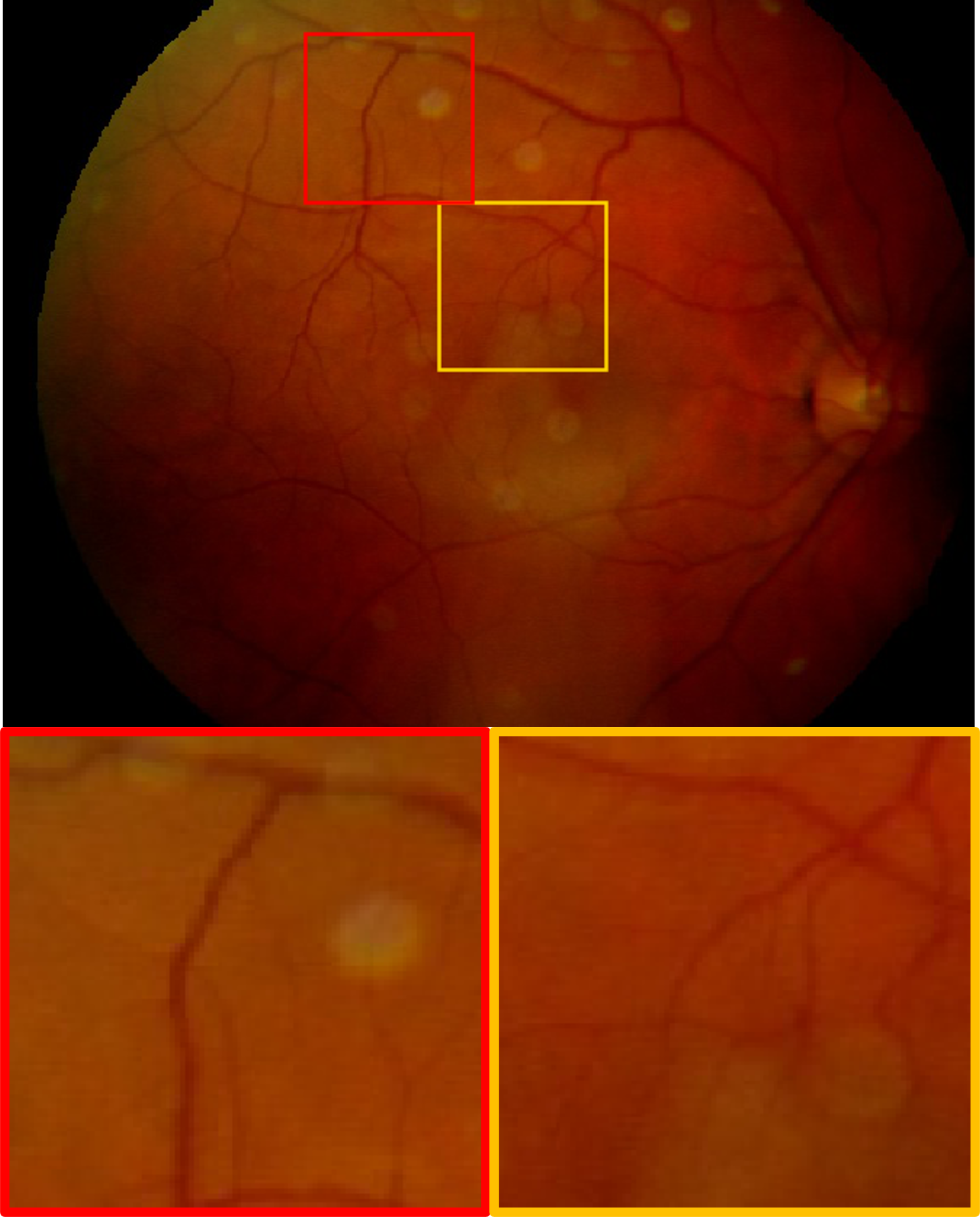} &
         \includegraphics[width=\imwidth\linewidth]{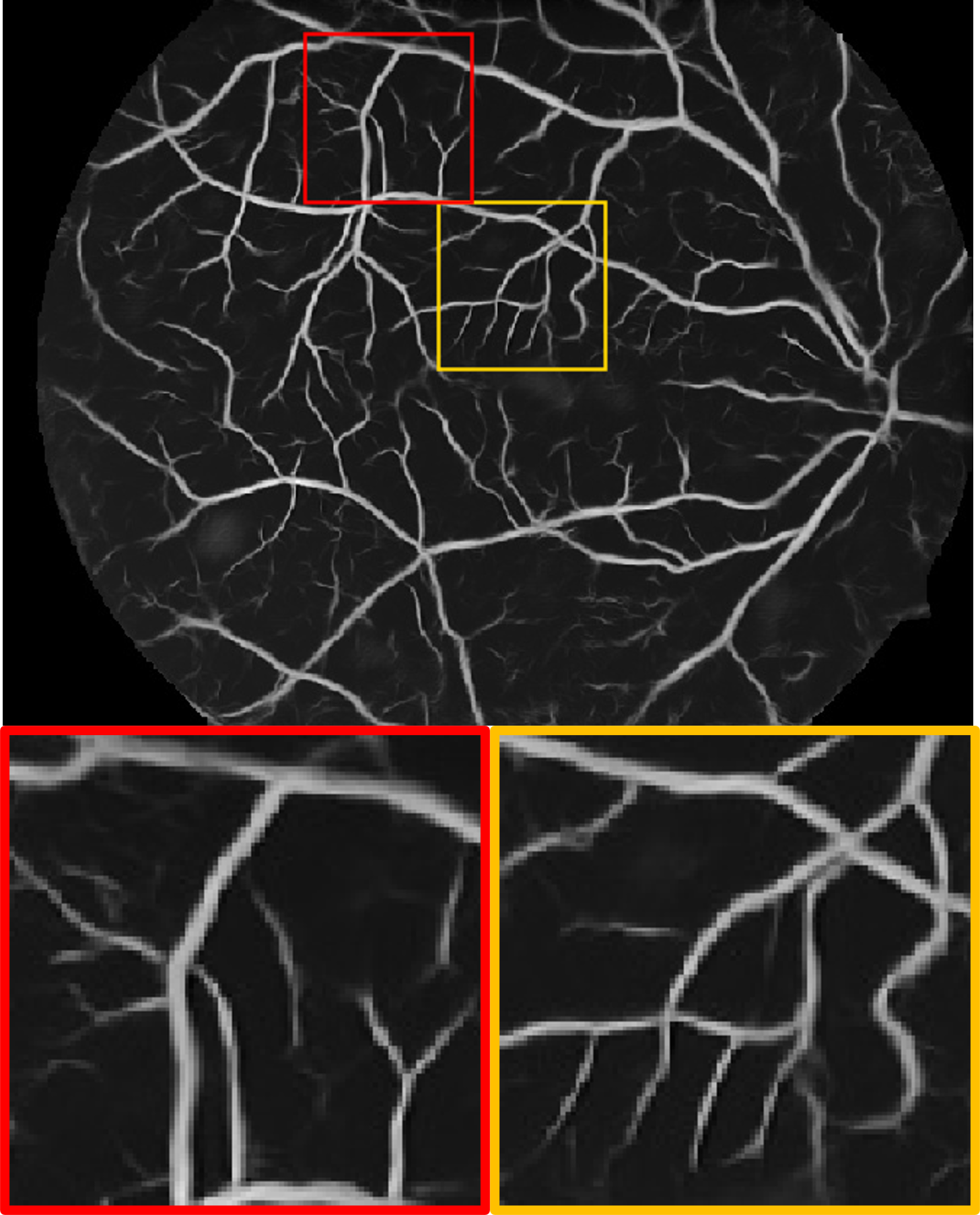} &
         \includegraphics[width=\imwidth\linewidth]{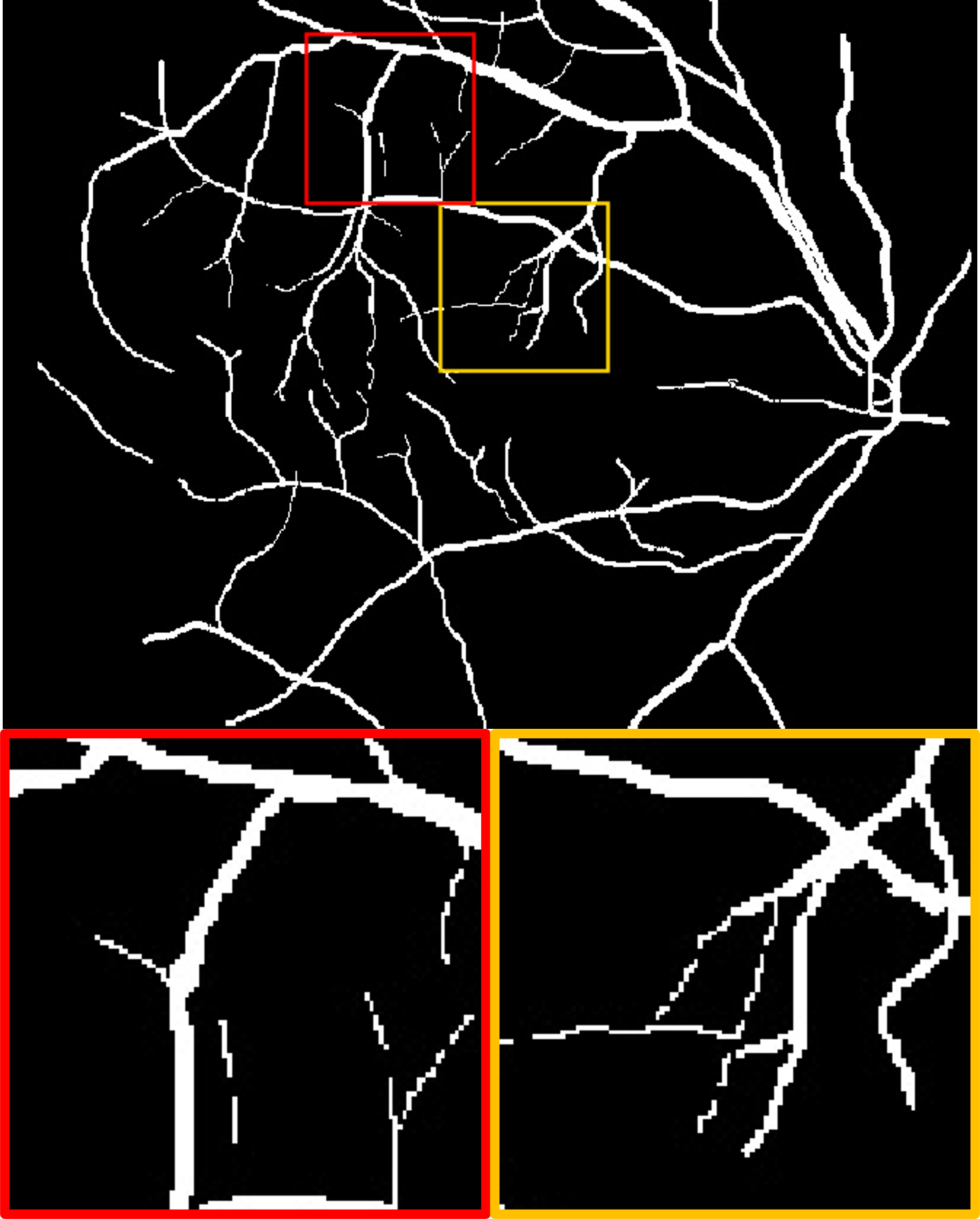} \\
         
         \rotatebox{90}{\hspace{3.0cm}\textbf{STARE}}
         \includegraphics[width=\imwidth\linewidth]{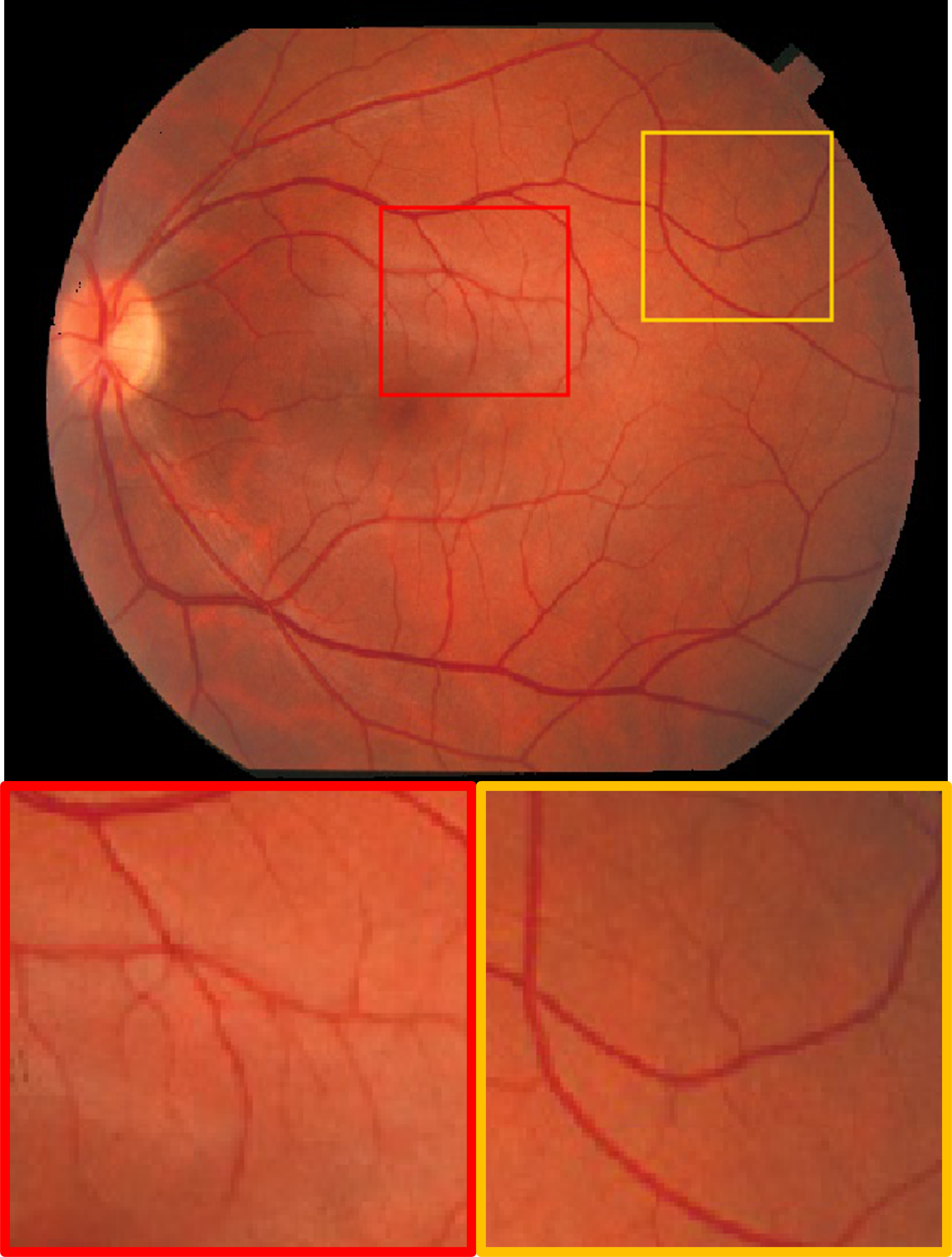} &
         \includegraphics[width=\imwidth\linewidth]{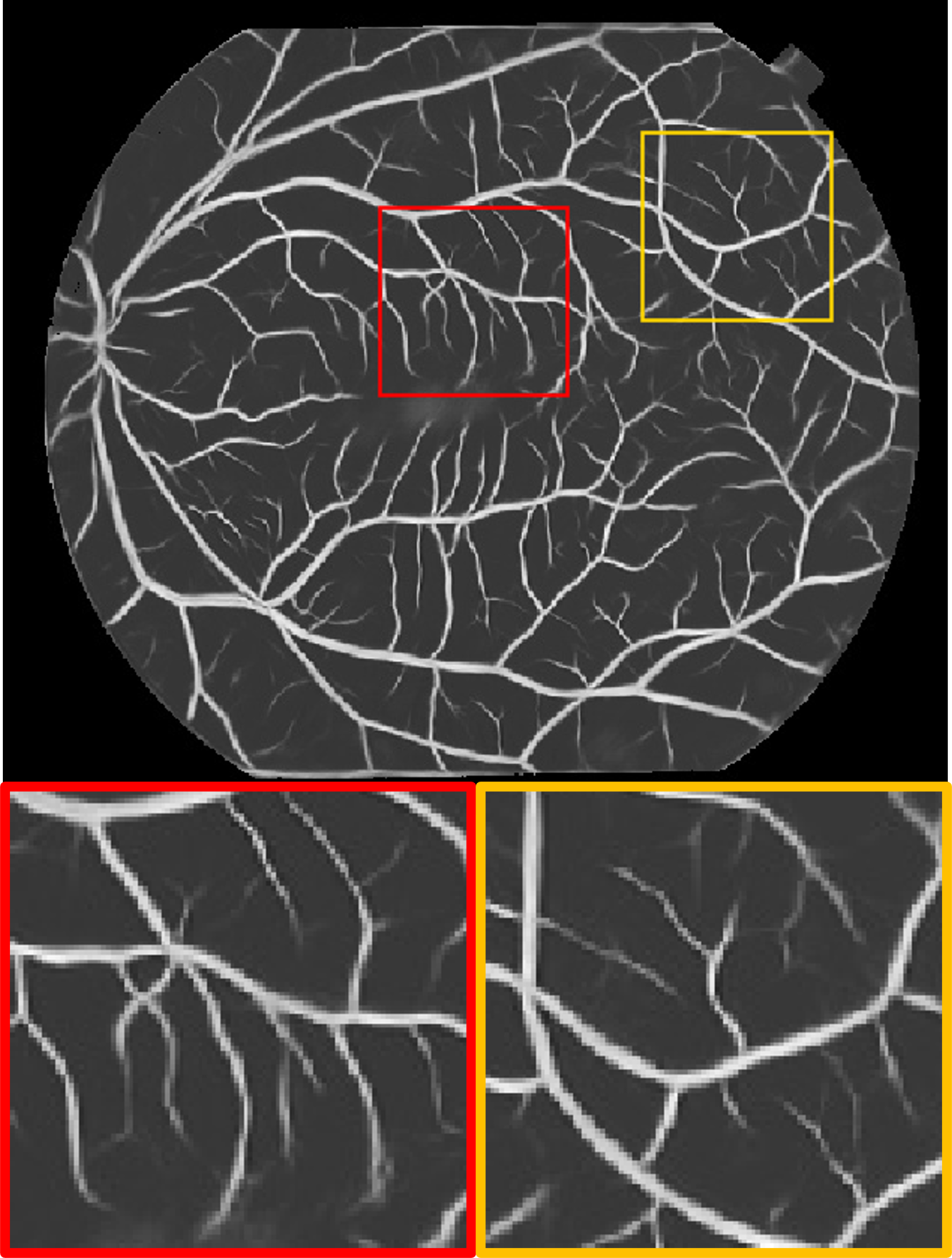} &
         \includegraphics[width=\imwidth\linewidth]{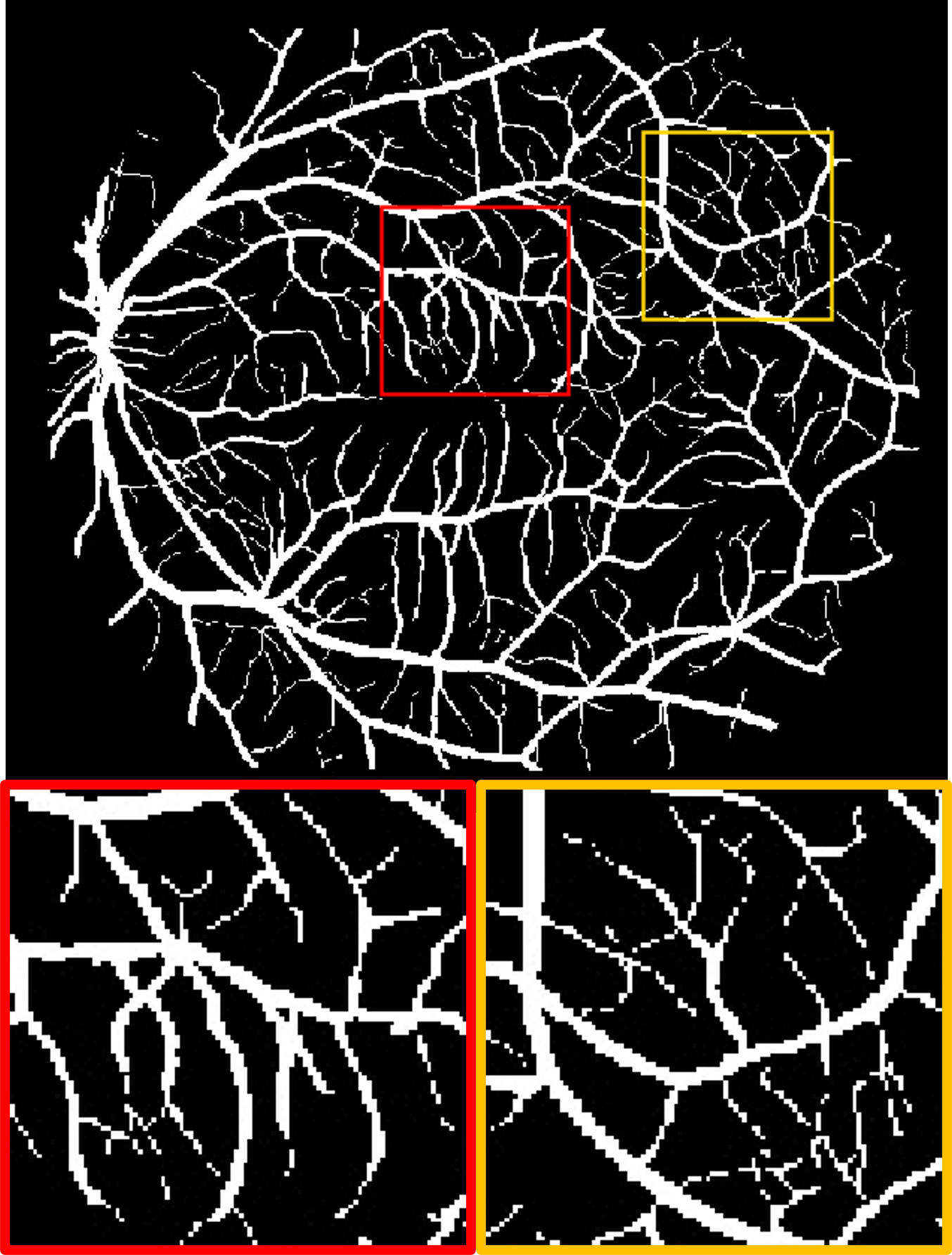} \\
    \end{tabular}
    \caption{Test examples from the target domains, ARIA and STARE. Below each image, close-up panels show two highlighted areas (red and yellow boxes) for easier comparison. Deep angiograms provide excellent vessel clarity. }
    \label{fig:result_image}
\end{figure}

\subsection{Domain Randomization}
There are two major causes for distribution shift of fundus images. First, within a well-curated dataset (e.g., DRIVE~\cite{staal2004ridge}), the image contrast is usually consistent. A model trained on such a dataset may struggle with a poor-contrast test image. Second, since a given dataset is unlikely to exhaustively provide samples of all possible pathologies, unseen features such as drusen and hemorrhages can be problematic during testing. 

To improve the robustness of the model, we randomize the source domain data by CLAHE~\cite{reza2004realization} in addition to other commonly used augmentation methods (e.g., rotation). For an input image $x$, we apply CLAHE $C_{\epsilon}$ to all the color channels with a random clip limit $\epsilon\in \mathcal{N}(5,1)$. In the resultant image $x'$, the contrast of vessels are strongly enhanced, as well as the background noise. Then as in Fig.~\ref{fig:model}, we introduce a contrastive loss $\mathcal{L}_{cont}$ for the latent image to guarantee that the model is not distracted by this exaggerated noise and provides stable visualization for input with various contrasts. The loss function is defined as the sum of the $L_2$ loss and the structural similarity (SSIM) loss.
\begin{equation}
    \mathcal{L}_{cont}=\|E_{\theta}(x)-E_{\theta}(x')\|_2 + SSIM(E_{\theta}(x)-E_{\theta}(x'))
\end{equation}

The SSIM loss is defined as
$    SSIM(x,y)=\frac{(2\mu_x\mu_y+c_1)(2\sigma_{xy}+c_2)}{(\mu_x^2+\mu_y^2+c_1)(\sigma_x^2+\sigma_y^2+c_2)}$,
where $\mu$ and $\sigma$ represent the mean and standard deviation of the image, and $c_1$ and $c_2$ are constants.

\subsection{Experiments}
\textbf{Baseline Methods.}
Since the color image is more sensitive to domain shift, it is common to convert the fundus image to grayscale as pre-processing, typically by extracting the green channel or using principle component analysis (PCA). We train a segmentation network that has the same architecture as $E_{\theta}$ with either the green channel or the PCA as input. We compare these two networks to Otsu thresholding of deep angiograms. 

\textbf{Datasets.} 
We use four publicly available fundus datasets as shown in Fig.~\ref{fig:model}(b). The {\bf \underline{DRIVE}} dataset~\cite{staal2004ridge} consists of 20 labelled images of size $565\times 584$. The {\bf\underline{HRF}} dataset~\cite{budai2013robust} contains 45 labelled images of size $3504\times 2336$. The {\bf\underline{STARE}} dataset~\cite{hoover2000locating} includes 20 labelled images of size $700\times 605$.
The {\bf\underline{ARIA}} dataset~\cite{farnell2008enhancement} includes 138 labelled images of size $768\times 576$.
DRIVE and HRF are set as source domain, whereas STARE and ARIA are used for testing.

\textbf{Implementation Details.}
All networks are trained and tested on an NVIDIA RTX 2080TI 11GB GPU. We use a batch size of 4 and train for 300 epochs. We use the Adam optimizer with the initial learning rate of $5\times 10^{-4}$ for the proposed VAE, $1\times 10^{-3}$ for the baseline segmentation networks. The learning rate for both networks decay by 0.5 every 3 epochs.

\section{RESULTS and Conclusion}

\begin{figure}[t]
    \centering
    \begin{tabular}{c@{}c}
         \includegraphics[width=0.46\linewidth]{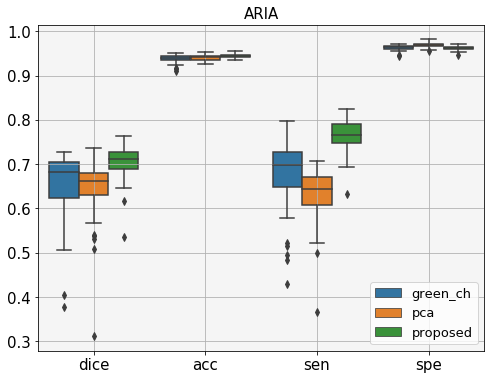} &  
         \includegraphics[width=0.46\linewidth]{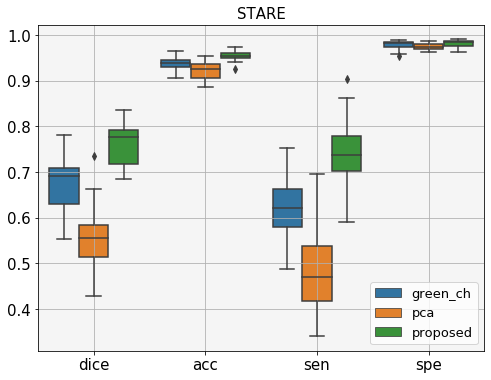} \\
    \end{tabular}
    \caption{Quantitative evaluation of the segmentation results on the two target domains. From left to right: Dice coefficient, accuracy, sensitivity, specificity. {\bf Blue,}  segmentation network trained on the green channel. {\bf Orange,} segmentation network trained on the PCA image. {\bf Green,} segmentation obtained by thresholding the deep angiogram.}
    \label{fig:boxplot}
\end{figure}

Fig.~\ref{fig:result_image} shows a test example from each of the target domains. We observe that for different datasets, the manual annotations includes varying amounts of detail: the label for the STARE dataset contains many more small vessels than ARIA. In the ARIA example, the deep angiogram is able to enhance the thin vessels with very poor contrast. This is also evident by the big vessels seen at the bottom left quadrant of the image where the illumination is low. Moreover, the angiogram filters out the circular artifacts seen within the red box. In the STARE example, our model extracts most of the vasculature including the faintly visible fine vessels. These tiny vessels  have relatively lower intensity in the deep angiogram, which suggests lower confidence. Compared to the manual label, the deep angiogram can also delineate the vessel diameter more precisely. 

We quantitatively evaluate the 
vessel segmentation performance in Fig.~\ref{fig:boxplot}. 
By simple thresholding on deep angiogram, we obtain get better vessel maps than the segmentation networks that use the green channel and PCA image as inputs.

The proposed method can effectively extract a specific type of feature from a complex context. Specific to retinal vessels, our model can generate stable deep angiograms that dramatically enhance small vessels with poor contrast for color fundus images from unseen domains. Hence, deep angiogram is a low-cost method that can be performed using  standard fundus photography technologies, including portable handheld systems. The ability to resolve  vascular features without the need for exogenous contrast injections significantly reduces the clinical expertise/equipment/cost of retinal angiography. Integration of these technologies with recent demonstrations of cellphone-based fundus photography methods and remote diagnostic technologies can move retinal disease screening out of the clinic and dramatically expand the impact of color fundus photography.

\section{ACKNOWLEDGEMENTS}
This work is supported by the Vanderbilt University Discovery Grant Program.

\bibliography{report} 
\bibliographystyle{spiebib} 

\end{document}